\documentclass[aps,prc,showpacs,nofootinbib]{revtex4}

\usepackage{epsfig}
\usepackage{graphicx}

\begin{document}
\title{Nucleon electromagnetic form factors and polarization observables in space-like and time-like regions}
\author{Egle Tomasi-Gustafsson, F. Lacroix, C. Duterte and G. I. Gakh\footnote{Permanent address:\it National Science Center KFTI, 310108 Kharkov, Ukraine}}
\affiliation{\it DAPNIA/SPhN, CEA/Saclay, 91191 Gif-sur-Yvette Cedex,
France}
\date{\today}

\begin{abstract}
We perform a global analysis of the experimental data of the electromagnetic nucleon form factors, in space-like and time-like regions. We give the expressions of the observables in annihilation processes, such as $p+\overline{p}\to \ell^+ +\ell^- $, $\ell=e$ or $\mu$, in terms of form factors. We discuss some of the phenomenological models proposed in the literature for the space-like region, and consider their analytical continuation to the time-like region. After determining the parameters through a fit on the available data, we give predictions for the observables which will be experimentally accessible with large statistics, polarized annihilation reactions.
\end{abstract}
\maketitle

\section{Introduction}

Elastic hadron electromagnetic form factors (FFs) are fundamental quantities for the understanding of nucleon structure. They contain information on the nucleon ground state, and constitute a further severe test for the models of nucleon structure, which already reproduce the static properties of the nucleon, such as masses and magnetic moments. Moreover, the dependence of FFs on the momentum transfer squared, $q^2=-Q^2$, should reflect the transition from the non perturbative regime, where effective degrees of freedom describe the nucleon structure, to the asymptotic region, where QCD applies.

The magnetic proton FF, which is the dominant term in the elastic $ep$ cross section, has been measured at $Q^2$ values up to 31 GeV$^2$ in the space-like (SL) region \cite{Ar86}, and from $p \bar p$ or $e^+e^-$ annihilation up to 18 GeV$^2$  in the time-like (TL) region \cite{An03}. Large progress has been recently done in the determination of the electric and magnetic proton form factors, based on the idea, firstly suggested in Ref. \cite{Re68}, to measure the polarization of the recoil proton in $\vec e p$ elastic scattering, when the electron is longitudinally polarized. 

Experiments, based on this method, have been performed at JLab up to $Q^2=5.6$ GeV$^2$ \cite{Jo00,Ga02}. A similar method, applied to the reaction $d(e,e'n)$p in quasi-elastic kinematics, has allowed the measurement of the neutron electric FF up to $Q^2$=1 GeV$^2$ using a polarized deuteron target \cite{Day} and up to $ Q^2$=1.47 GeV$^2$, measuring the polarization of the outgoing neutron \cite{Madey}. The polarization method has been also successfully applied at low $Q^2$, for a precise determination of the neutron FFs, at Mainz, and shows that $G_{En}$ is definitely different from zero (\cite{Gl04} and refs therein). 

These results have been obtained thanks to the availability of high intensity, highly polarized electron beams and polarized targets, and to the optimization of hadron polarimeters in the GeV range.  An extension of the measurement of the polarization transfer in $\vec e +p\to e+\vec p$ up to 9 GeV$^2$ is in preparation \cite{04108}.

More data are expected in future, in SL region, after the upgrade of Jlab, and in TL region, at Frascati and at the future FAIR facility at Darmstadt \cite{GSI}.

In the TL region \cite{An03}, due to the poor statistics, the determination of FFs requires to integrate the differential cross section over a wide angular 
range. One typically assumes that the $G_E$ contribution plays a minor role in the cross section at large $q^2$ and the 
experimental results are usually given 
in terms of $|G_M|$, under the hypothesis that $G_E=0$ or $|G_E|=|G_M|$. The first hypothesis is an arbitrary one. The second hypothesis is strictly 
valid at threshold only, i.e., for $\tau=q^2/(4m^2)=1$, but there is no 
theoretical argument which justifies its validity at any other momentum 
transfer, where $q^2\neq 4m^2$ ($m$ is the nucleon mass).

The measurement of the differential 
cross section for the process $p+\overline{p}\to \ell^+ +\ell^-$ at a fixed value of the total energy $s$, and for two different angles $\theta$,  allowing  the separation of the two FFs, $|G_M|^2$ and $|G_E|^2$, is equivalent to the well known Rosenbluth separation for the elastic $ep$-scattering. However, in TL region, this procedure is simpler, as it requires to change only one kinematical variable, $\cos\theta$, whereas, in SL region 
it is necessary to change simultaneously two kinematical variables: the energy of the initial electron and the electron scattering angle, fixing the momentum transfer squared, $Q^2$. Due to the limited statistics, the Rosenbluth separation of the $|G_E|^2$ and $|G_M|^2$ contributions has not yet been realized in TL region. Early attempts showed that the large error bars prevent to discriminate between the two hypothesis on  $|G_E|$ and $|G_M|$ quoted above  \cite{Bi83,Ba94}. 

The $|G_M|$ values depend, in principle, on the kinematics where the 
measurement was performed and the angular 
range of integration. However, it turns out that these two assumptions for $G_E$ 
lead to comparable values for $|G_M|$.

In the SL region the situation is different. The cross section for the elastic 
scattering 
of electrons on protons is sufficiently large to allow the measurements of the 
angular 
distribution and/or of polarization observables. Data on $G_M$ are available up to the highest 
measured value, $Q^2\simeq$ 31 GeV$^2$ \cite{Ar86} and this FF is often approximated according to a dipole behavior:
\begin{equation}
G_M(Q^2)/\mu_p=G_d,~\mbox{with}~
G_d=\left [1+{Q^2}/{ m_d^2 }\right ]^{-2},~m_d^2=0.71~\mbox{GeV}^2, 
\label{eq:dipole}
\end{equation}
where $\mu_p$ is the magnetic moment of the proton.

It should be noted that the independent determination of both FFs, $G_M$ and $G_E$, from the unpolarized $e^- +p$-cross section, has been done up to $Q^2=$ 8.7 GeV$^2$ \cite{And94}, and the further extraction of $G_M$  assumes $G_E=G_M/\mu_p$.
The behavior of $G_{Ep}$, deduced from polarization experiments, in which, more precisely, the ratio $G_{Ep}/G_{Mp}$ is directly related to the longitudinal and transversal component of the scattered proton polarization, differs from $G_M/\mu_p$,
with a deviation up to 70\% at $Q^2$=5.6 GeV$^2$ \cite{Ga02}. This is the maximum momentum at which new, precise data are available.

The recent experimental data have inspired many new theoretical 
developments, and shown the necessity of a global representation of FFs in the full region of momentum transfer squared. 

FFs are analytical functions of $q^2$, 
being real functions in the SL  region (due to the hermiticity of the 
electromagnetic Hamiltonian) and complex functions in the 
TL region. The discussion of the constraints and consequences of a description in the full kinematical domain was firstly done in Ref. \cite{Bi93} and more recently in Refs. \cite{ETG01,Br03,Ia03,Wa04}.

The extension of the nucleon models developed for the SL region to the TL region is straightforward for VMD inspired models, which may give a good description of all FFs in the whole kinematical region, after a fitting procedure involving a certain number of parameters \cite{Du03,Bij04}. 

The purpose of this paper is to update and compare some of the available models  on the world data set in both TL and SL regions, and to predict time-like polarization observables, in framework of these models. The paper is organized as follows. In section II the expressions for the relevant polarization observables, in the process $\bar p+p\to  \ell^+ +\ell^- $, $\ell=e$ or $\mu$,  are given as a function of the electromagnetic FFs, in Section III we update some of the fits of nucleon FFs on the available data, and discuss their extension to the TL region. In section IV we give the predictions of the considered models in TL region.

\section{Observables in TL region}
We develop a simple and transparent formalism for the study of polarization phenomena for 
$p+\overline{p}\to \ell^+ +\ell^-$, in framework of one-photon mechanism.  

\begin{center}
\begin{figure}[ht]
\mbox{\epsfxsize=8cm\leavevmode\epsffile{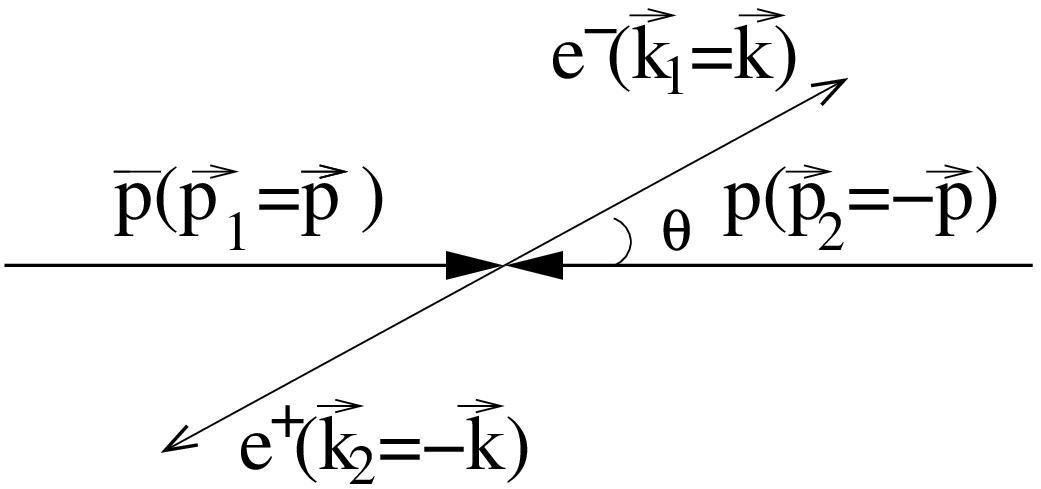}}
\caption{The kinematics of the process $p+\overline{p}\to e^- + e^+$ in the reaction CMS.}
\label{fig:cms}
\end{figure}
\end{center}

The calculations of the cross section and of the polarization observables for the process $\bar p+p\to  \ell^+ +\ell^- $, $\ell=e$ or $\mu$, in the annihilation channel are more conveniently performed in the center of mass  system (CMS), Fig. \ref{fig:cms}. The momenta of the particles are indicated in the figure and  $|\vec k_1|=|\vec k_2| =|\vec k|$ and $|\vec p_1|=|\vec p_2| =|\vec p|$. Let us choose the $z$ axis along the direction of the incoming antiproton, the $y$ axis normal to the scattering plane, and the $x$ axis to form a left-handed coordinate system. The components of the unity vectors are therefore $\hat{\vec p}=(0,0,1)$ and $\hat{\vec k}=(\sin\theta,0,\cos\theta)$ with $\hat{\vec p}\cdot\hat{\vec k}=\cos\theta$, where $\theta$ is the electron production angle in CMS. The relevant kinematical variable is the antiproton energy, $E$, which is related to the four momentum transfer, $q^2=s=(k_1+k_2)^2=4E^2$, as, in CMS, $\vec k_1+\vec k_2=0$. In the laboratory (Lab) system, one finds $q^2=2m^2+2mE_L$, where $E_L$ is the Lab antiproton energy. The observables are calculated in the approximation of zero electron mass.

The starting point of the analysis of the reaction $p+\overline{p}\to e^+ +e-$ is the standard expression of the matrix element in framework of one-photon exchange mechanism:
\begin{equation}
{\cal  M}=\displaystyle\frac {e^2}{q^2}\overline{u}(-k_2)\gamma_{\mu}u(k_1) \overline{u}(p_2)\left [F_{1N}(q^2)\gamma_{\mu}-
\displaystyle\frac{\sigma_{\mu\nu}q_{\nu}}{2m}F_{2N}(q^2)\right] u(-p_1),
\label{eq:mat}
\end{equation}
where  $p_1$, $p_2$, $k_1$ and $k_2$  are the four-momenta of initial antiproton and proton and  the final electron and positron respectively,   $q^2>4m^2$, $q=k_1+k_2=p_1+p_2$. $F_{1N}$ and $F_{2N}$ are the Dirac and Pauli nucleon electromagnetic FFs, which are complex functions of the variable $q^2$ - in the TL region of momentum transfer. 
In framework of one-photon exchange, the matrix element is written as the product of the leptonic and hadronic currents:
\begin{equation}
{\cal M}=\displaystyle\frac{e^2}{q^2} \ell_{\mu}{\cal J}_{\mu}=
\displaystyle\frac{e^2}{q^2} (\ell_0{\cal J}_0- \vec\ell\cdot\vec{\cal J})
=-\displaystyle\frac{e^2}{q^2} \vec\ell\cdot\vec{\cal J},
\label{eq:eq1}
\end{equation}
where $\ell_0{\cal J}_0=0$, due to the conservation of the leptonic and hadronic currents\footnote{The conservation of the current implies that $\ell\cdot q=0$, i.e., $\ell_0 q_0-\vec\ell\cdot\vec q =0$, but $\vec q=\vec k_1+\vec k_2=0 $ in CMS. Therefore, $\ell_0 q_0=0$ for any energy $q_0$, i.e., $\ell _0=0.$}. The expression for the leptonic current is:
\begin{equation}
\vec\ell=\sqrt{q^2}\phi^{\dagger}_2(\vec\sigma-\hat{\vec k}\vec\sigma\cdot\hat{\vec k})\phi_1,
\label{eq:eq2}
\end{equation}
where $\phi_1(\phi_2)$ is the two-component spinor of the electron (positron), $\hat{\vec k}$ is the unit vector along the final electron three-momentum, and for the hadronic current:
\begin{equation}
\vec{\cal J}=\sqrt{q^2}\chi^{\dagger}_2\left [ G_M(q^2)(\vec\sigma-
\hat{\vec p}\vec\sigma\cdot\hat{\vec p})+\displaystyle\frac{1}{\sqrt\tau}G_E(q^2)\hat{\vec p}\vec\sigma\cdot\hat{\vec p} \right ]\chi_1, 
\label{eq:eq3}
\end{equation}
where $\chi_1$ and $\chi_2$ are the two-component spinors of the antiproton and the  proton, $\hat{\vec p} $ is the unit vector along the three momentum of the antiproton in CMS. 
\begin{center}
\begin{figure}
\mbox{\epsfxsize=9.cm\leavevmode\epsffile{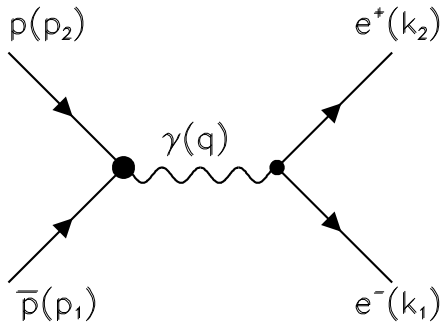}}
\caption{One-photon mechanism for $p+\overline{p}\to e^- + e^+$ (with notation of four particle four-momenta).}
\label{fig:borns}
\end{figure}
\end{center}
From this expression one can see the physical meaning of the  particular relation between the nucleon electromagnetic FFs at threshold:
$$
 G_{E}(q^2)=G_{M}(q^2),~q^2= 4 m^2.
$$ 
The structure $\hat{\vec p}\vec\sigma\cdot\hat{\vec p}$  describes the $\overline{p}
+p$ annihilation from $D$-wave, i.e.,  with angular momentum $\ell$=2. At threshold, where $\tau\to 1$, the finite radius of the strong interaction allows only the S-state, and  $G_{M}(q^2)-\displaystyle\frac{1}{\sqrt\tau}G_{E}(q^2)=0$.

From Eqs. (\ref{eq:eq1}), (\ref{eq:eq2}), and (\ref{eq:eq3}) one can find the formulas for the unpolarized cross section, the angular asymmetry and all the polarization observables.

\subsection{The cross section}

To calculate the cross section when all particles are unpolarized, one has to sum over the polarization of the final particles and to average over the polarization of initial particles:
$$
\left (\displaystyle\frac{d\sigma}{d\Omega}\right )_0=\displaystyle\frac{| \overline{\cal  M}|^2}{64\pi^2 q^2}
\displaystyle\frac{k}{p},~ k=\displaystyle\frac{\sqrt{(q^2)}}{2},~p=\sqrt{\displaystyle\frac{(q^2)}{4}-m^2},
$$
\begin{equation}
| \overline{\cal  M}|^2=\displaystyle\frac{1}{4}\displaystyle\frac{e^4}{q^4}
\ell_{ab} {\cal J}_{ab},~\ell_{ab}=\ell_a\ell_b^*,~
{\cal J}_{ab}={\cal J}_a{\cal J}_b^*.
\label{eq:eq12}
\end{equation}
Using the expressions (\ref{eq:eq2}) and (\ref{eq:eq3}), the formula for the cross section in CMS is:
\begin{equation}
\left (\displaystyle\frac{d\sigma}{d\Omega}\right )_0={\cal N}\left [(1+\cos^2\theta)|G_M|^2+\displaystyle\frac{1}{\tau}\sin^2\theta|G_E|^2\right ],
\label{eq:eq7}
\end{equation}
where 
${\cal N}=\displaystyle\frac{\alpha^2}{4\sqrt{q^2(q^2-4m^2)}}$, $\alpha=e^2/(4\pi)\simeq 1/137 $, is a kinematical factor. This formula was firstly obtained in Ref. \cite{Zi62}.

The angular dependence of the cross section, Eq. (\ref{eq:eq7}), results 
directly from the assumption of one-photon exchange, where the photon has spin 1 and the electromagnetic hadron interaction satisfies the 
$P-$invariance. 
Therefore, the measurement of the differential 
cross section at three angles (or more) would also allow to test the presence of $2\gamma$ exchange \cite{Re03}. 

The electric and the magnetic FFs are weighted by different angular terms, in the cross section, Eq. (\ref{eq:eq7}). One can define an angular asymmetry, ${\cal R}$, with respect to the differential cross section measured at $\theta=\pi/2$, $\sigma_0$ \cite{ETG01}:
\begin{equation}
\left (\displaystyle\frac{d\sigma}{d\Omega}\right )_0=
\sigma_0\left [ 1+{\cal R} \cos^2\theta \right ],
\label{eq:asym}
\end{equation}
where ${\cal R}$ can be expressed as a function of FFs:
\begin{equation}
{\cal R}=\displaystyle\frac{\tau|G_M|^2-|G_E|^2}{\tau|G_M|^2+|G_E|^2}.
\end{equation}
This observable should be very sensitive to the different underlying 
assumptions on FFs, therefore, a precise measurement of this quantity, which does not require polarized particles, would be very  interesting. 

The $q^2$ dependence of the total cross section can be presented as follows:
\begin{equation}
\sigma(q^2)={\cal N}\displaystyle\frac{8}{3}\pi \left [2|G_M|^2+ \displaystyle\frac{1}{\tau}|G_E|^2\right ]. 
\label{eq:eq13}
\end{equation}

Polarization phenomena will be especially important in $p+\overline{p}\to \ell^+ +\ell^-$. The dependence of the cross section on the polarizations $\vec P_1$ and $\vec P_2$ of the colliding antiproton and proton can be written as follows:
\begin{eqnarray}
\left (\displaystyle\frac{d\sigma}{d\Omega}\right )_0(\vec P_1,\vec P_2)
&=& \left (\displaystyle\frac{d\sigma}{d\Omega}\right )_0
[1+A_y(P_{1y}+ P_{2y})+A_{xx} P_{1x}P_{2x}+A_{yy} P_{1y}P_{2y}+A_{zz} P_{1z}P_{2z}\nonumber \\
&&
+A_{xz} (P_{1x}P_{2z}+P_{1z}P_{2x})],
\label{eq:eq13a}
\end{eqnarray}
where the coefficients $A_i$ and $A_{ij}$ $(i,j=x,y,z)$, analyzing powers and correlation coefficients, depend on the nucleon FFs. Their  explicit form is given in the following sections. The dependence (\ref{eq:eq13a}) results from the P-invariance of hadron electrodynamics.

\subsection{Single spin polarization observables}

In case of polarized antiproton beam with polarization $\vec P_1$, the contribution to the cross section can be calculated as:
\begin{equation}
\left (\displaystyle\frac{d\sigma}{d\Omega}\right )_0 \vec A_1=-\ell_{ab}\displaystyle\frac{1}{4} Tr {\cal J}_a\vec\sigma {\cal J}_b^*.
\label{eq:eq14}
\end{equation}
Here the terms related to $|G_E|^2$ and $|G_M|^2$ vanish. For the interference terms, the only non zero analyzing power is related to $P_y$:
\begin{equation}
\left (\displaystyle\frac{d\sigma}{d\Omega}\right )_0 A_{1,y}=\displaystyle\frac{\cal N}{\sqrt{\tau}}\sin2\theta Im(G_MG_E^*).
\label{eq:eq15}
\end{equation}
When the target is polarized, one writes:
$$
\left (\displaystyle\frac{d\sigma}{d\Omega}\right )_0\vec A_2=\ell_{ab}\displaystyle\frac{1}{4} Tr {\cal J}_a{\cal J}_b^*
\vec\sigma. $$
Again the terms related to $|G_E|^2$ and $|G_M|^2$ vanish. Moreover, one can find $\vec A_2=\vec A_1=\vec A$.

Eq. (\ref{eq:eq15}) has been proved also in Ref. \cite{Zi62}. One can see that this analyzing power, being T-odd, does not vanish in $p+\overline{p}\to \ell^+ +\ell^-$, even in one-photon approximation, due to the fact FFs are complex in time-like region. This is a principal difference with elastic $ep$ scattering. Let us note also that the assumption $G_E=G_M$ implies $A_y=0$, independently from any model taken for the calculation of FFs.

\subsection{Double spin polarization observables}

The contribution to  the cross section, when both colliding particles are polarized is calculated through the following expression:
$$
\left (\displaystyle\frac{d\sigma}{d\Omega}\right )_0
A_{ab}=-\displaystyle\frac{1}{4} \ell_{mn} Tr {\cal J}_m\sigma_a{\cal J}_n^{\dagger}\sigma_b, 
$$
where $a$ and $b=x,y,z$ refer to the $a(b)$ component of the projectile (target) polarization. Among the nine possible terms, $A_{xy}=A_{yx}=A_{zy}=A_{yz}=0$, and the nonzero components are:
\begin{eqnarray}
\left (\displaystyle\frac{d\sigma}{d\Omega}\right )_0A_{xx}&=& 
\sin^2\theta\left (|G_M|^2 +\displaystyle\frac{1}{\tau}|G_E|^2\right ){\cal N},\nonumber \\
\left (\displaystyle\frac{d\sigma}{d\Omega}\right )_0A_{yy}&=& 
-\sin^2\theta\left (|G_M|^2 -\displaystyle\frac{1}{\tau}|G_E|^2\right ){\cal N},\nonumber\\
\left (\displaystyle\frac{d\sigma}{d\Omega}\right )_0A_{zz}&=& 
\left [(1+\cos^2\theta)|G_M|^2-
\displaystyle\frac{1}{\tau}\sin^2\theta |G_E|^2\right ]{\cal N},\nonumber\\
\left (\displaystyle\frac{d\sigma}{d\Omega}\right )_0A_{xz}&=&\left (\displaystyle\frac{d\sigma}{d\Omega}\right )_0A_{zx}=
\displaystyle\frac{1}{\sqrt{\tau}}\sin 2\theta Re G_E G_M^* {\cal N}. \label{eq:pol}
\end{eqnarray}
One can see that the double spin observables depend on the moduli squared of FFs, besides $A_{xz}$. Therefore, in order to determine the relative phase of FFs, in TL region, the interesting observables are $A_y$, and $A_{xz}$ which contain, respectively, the imaginary and the real part of the  product $G_EG_M^*$.

\section{Results and discussion}

\subsection{The data}
The nucleon FFs world data were collected and listed in Table I for proton FFs and Table II for neutron FFs in SL region.

\begin{figure}[pht]
\begin{center}
\includegraphics[width=16cm]{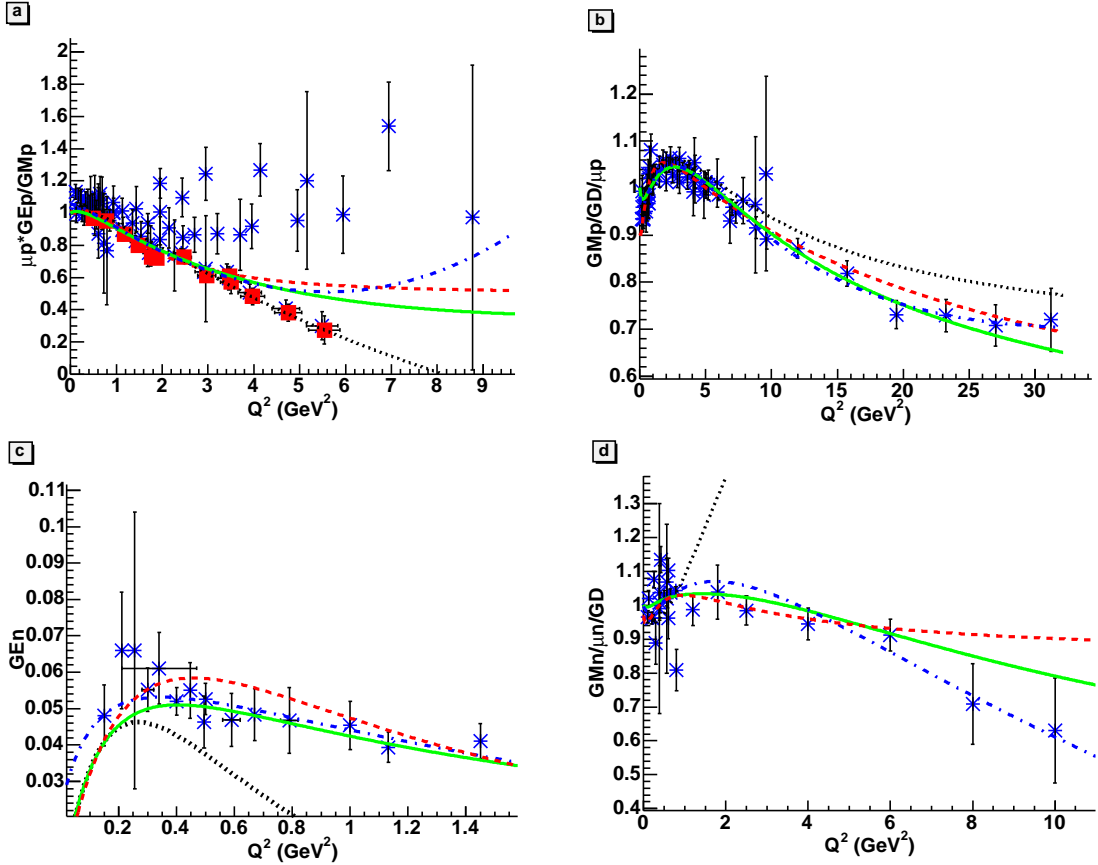}
\caption{\label{fig:fig1} Nucleon Form Factors in Space-Like region: (a) proton electric FF, scaled by $\mu_p G_{Mp}$ (b) proton magnetic FF scaled by $\mu_p G_D$ , (c) neutron electric FF, (d) neutron magnetic FF, scaled by $\mu_n G_D$. The predictions of the models are drawn:  from Ref. \cite{Ia73} (dotted line),  from Ref. \cite{Lomon} (solid line), model from Ref. Ref. \cite{Ho76} (dash-dotted line), from \cite{Bo95} (dashed line). }
\end{center}
\end{figure}

In Fig. \ref{fig:fig1} the nucleon FFs world data in SL region are shown: the ratio $\mu_p G_{Ep}/G_{Mp} $ (Fig. \ref{fig:fig1}a) , the magnetic proton FF normalized to the dipole FF and divided by $\mu_p$ (Fig. \ref{fig:fig1}b), the electric  neutron FF (Fig. \ref{fig:fig1}c), and the magnetic neutron FF normalized to the dipole FF and divided by $\mu_n$ (Fig. \ref{fig:fig1}d). 
For the electric proton FF, the discrepancy among the data measured with the Rosenbluth methods (stars) and the polarization method (solid squares) appears clearly in Fig. \ref{fig:fig1}a. This problem has widely been discussed in the literature, (for a recent discussion see, for instance, Ref. \cite{Pu05}) and rises fundamental issues. If the trend indicated by polarization measurements is confirmed at higher $Q^2$ \cite{04108}, not only the electric and magnetic charge distribution in the nucleus are different and deviate, classically, from an exponential charge distribution, but also the electric FF has a zero and becomes eventually negative. This scenario will change our view on the nucleon structure and will favor VMD inspired models like \cite{Du03,Bij04}, which can reproduce such behavior.

We included data issued from both kind of measurements in the fit, although if a consensus seems to appear that FFs extracted from polarization measurements are more reliable, as less affected by all kinds of radiative corrections. Our purpose here is not to get the best $\chi^2$, but to get a global description of the overall data. The precision and the number of points is very different for the different FFs, therefore one can obtain a good $\chi^2$ for a model that reproduces well, for example, the electric and magnetic FFs in the SL region and fails in giving the trend of $G_{En}$ in TL region.

We included in the fit the data on proton magnetic FFs which were published  after 1973, and we did not include the data on the neutron electric FFs from Refs.  \cite{Ha73,St66,Br95,Hu65,Ak64} as data of much better precision were, later, available, in the same $Q^2$ range. 

The data in the TL region are drawn in 
In Fig. \ref{fig:fig2}a, b  for the proton and in Fig. \ref{fig:fig2}c, d for the neutron, respectively and summarized in Table III. As no separation has been done for electric and magnetic FFs, the data are extracted under the hypothesis that $|G_{EN}|=|G_{MN}|$.
Concerning the neutron, the first and still unique measurement was done at Frascati, by the collaboration FENICE \cite{An98}.

\subsection{The models}

Among the existing models of nucleon FFs, we consider some parametrizations, which have an analytical expression that can be continued in TL region: predictions of pQCD, in a form generally used as simple fit to experimental data, a model based on vector meson dominance (VMD) \cite{Ia73}, and a third model based on an extension of VMD, with additional terms in order to satisfy the asymptotic predictions of QCD \cite{Lomon}, in the form called GKex(02L). We also considered the Hohler parametrization \cite{Ho76} and the Bosted empirical fit \cite{Bo95}.

In order to help the reader, we report in the Appendix the explicit forms of the parametrizations previously published, with the  parameters corresponding to the present fit, compared to the published ones.

\begin{figure}[pht]
\begin{center}
\includegraphics[width=16cm]{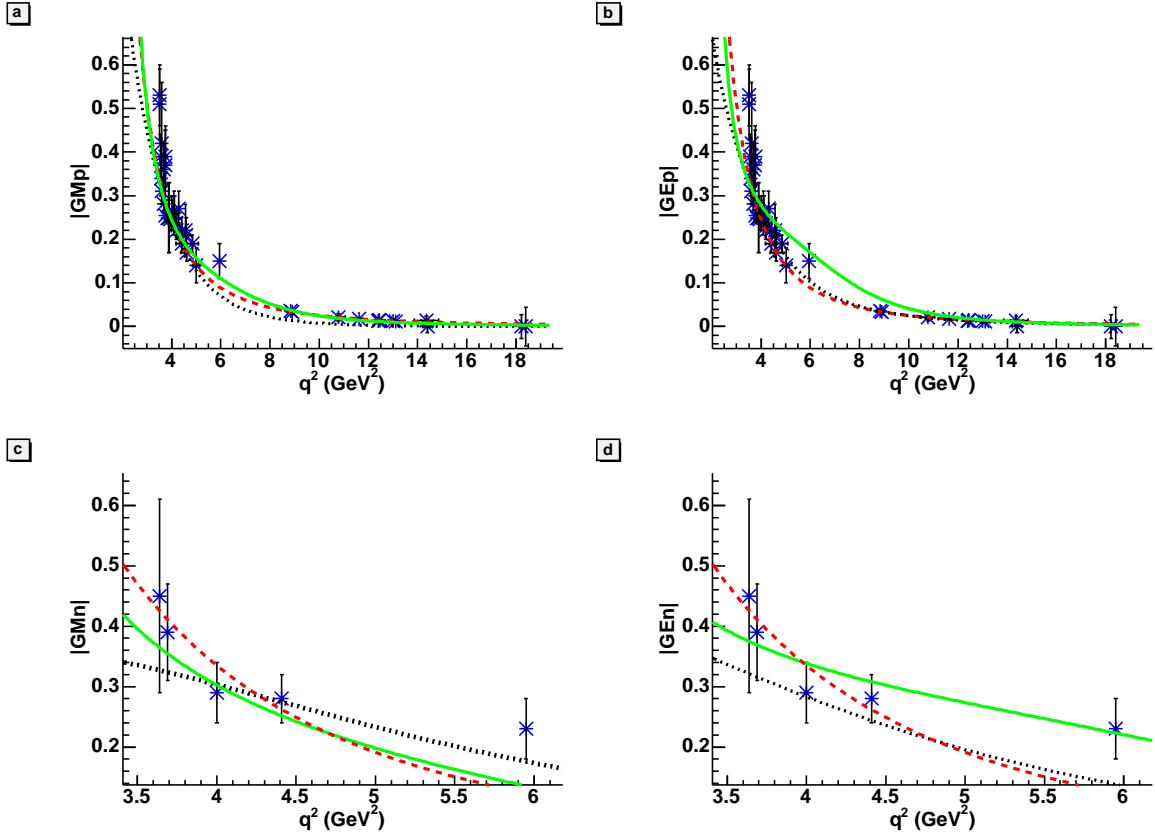}
\caption{\label{fig:fig2} Form Factors in Time-Like region and predictions of the models: pQCD-inspired  (dashed line), from Ref. \cite{Ia73} (dotted line), from Ref. \cite{Lomon} (solid line).}
\end{center}
\end{figure}

The pQCD prediction, based on counting rules, follows the dipole behavior (\ref{eq:dipole}) in SL region, and can be extended in TL region as \cite{Le80}:
\begin{equation}
|G_M|=\frac{A(N)}{q^4\ln^2(q^2/\Lambda^2)},
\label{eq:eqtp}
\end{equation}
where $\Lambda=0.3$ GeV is the QCD scale parameter and $A$ is a free parameter.
This simple parametrization is taken to be the same for proton and neutron. The best fit ( Fig. \ref{fig:fig2}, dashed line) is obtained with a parameter $A(p)$= 56.3 GeV$^4$ for proton and 
$A(n)$= 77.15 GeV$^4$ for neutron, which reflects the fact that in TL region, neutron FFs are larger than for proton. One should note that errors are also larger in TL region. 

A possible explanation of the fact that FFs are systematically larger in TL region than in SL region (which is true also in the proton case) is the presence of a resonance in the $N\overline{N}$ system, just below the $N\overline{N}$ threshold \cite{Ga96}.  

More pQCD inspired parametrizations exist for the form factor ratio $F_2/F_1$, which include logarithmic corrections, and have been recently discussed in Ref. \cite{Br03}. However, some of these analytical forms have problems related to the asymptotic behavior. This will be discussed in a future paper.

The analytical continuation to TL region of the other models is based on the following relations:
\begin{equation}
Q^2=-q^2=q^2e^{-i\pi}~\Longrightarrow~\left\{\begin{array}{c}
\ln(Q^2)=ln(q^2)-i\pi\\
\sqrt{Q^2}=e^{\frac{-i\pi}{2}}\sqrt{q^2}\\
\end{array} \right.
\end{equation}
Most of the models predict a different behavior for the electric and the magnetic FFs in TL region, whereas, as already mentioned,  no individual determination of electric and magnetic FFs has been done yet. We chose to fit the data assuming that they correspond to the magnetic FFs for proton and neutron, Fig. \ref{fig:fig2}a and \ref{fig:fig2}c, respectively. Therefore, the curves for the electric FFs, in Figs. \ref{fig:fig2}b and \ref{fig:fig2}d have to be considered predictions from the models. Including or not the data on neutron FFs, in TL region, influence very little the fitting procedure.

The parametrization from Ref. \cite{Ia73} is shown as a dotted line, in 
Figs. \ref{fig:fig1} and  \ref{fig:fig2}. This model is based on a view of the nucleon as composed by an inner core with a small radius (described by a dipole term) surrounded by a meson cloud.
While it reproduces very well the proton data in SL region (and particularly the polarization measurements), it fails in reproducing the large $Q^2$ behaviour of the magnetic neutron FF in SL region. The present fit constrained  on the TL data and on the recent SL data does not improve the situation. In framework of this model a good global fit in SL region has been obtained with a modification including a phase in the common dipole term. However, the TL region is less well reproduced \cite{Bij04}. Therefore, the curves drawn in all the figures correspond to the original parameters, which give, in our opinion, a better representation of the whole set of data. 

The result from an update fit based on the parametrization GKex(02L) \cite{Lomon} is shown in Figs. \ref{fig:fig1} and  \ref{fig:fig2} (solid line). It is possible to find a good overall parametrization, with parameters not far from those found in the original paper for the SL region only. The agreement is very good, for both proton and neutron FFs. 

The Hohler parametrization \cite{Ho76}, contains also pole terms with adjustable parameters. The $\rho$-exchange contribution, however, is fully determined, with constants fixed on $\pi N$ data. The model contains 17 parameters, already, so we did not try to readjust or refit the $\rho$-contribution. 
As noted in the original paper, such model is not suited to the extrapolation to TL region, because poles appear in the physical region. Constraining the parameters, in order to avoid these instabilities, worsens the description in the SL region.
Therefore, we give only a fit on all FFs, in  SL region, Fig. \ref{fig:fig1} (dash-dotted line), corresponding to $\chi^2/ndf\simeq 1.7$. The formulas as well as the original and updated parameters are also given in Appendix. Parametrization \cite{Du03} can be considered a successful generalization, in TL region, based on unitarity and analyticity. It requires the modelization of ten resonances, five isoscalar and five isovector.

The Bosted parametrization \cite{Bo95} is an empirical fit to nucleon FFs, in the SL region, based on simple formulas which are useful for fast estimations.  It does not seem possible to find a unique function, which describes satisfactorily both the magnetic nucleon FFs and the electric proton FF, so the parameters are specific to each FFs. In the extension to TL region, as for the Hohler parametrization, one can not avoid poles and instabilities, and attempts to obtain a description in SL and TL regions remained unsuccessful. Therefore, we give the fit for the SL region, only, as dashed line in Fig. \ref{fig:fig1}, and report in the Appendix the useful formulas and the updated parameters. As one can see from the table, they do not differ more than 20\% from the published ones and the fit corresponds to $\chi^2/ndf\simeq 2$.

\section{Predictions in TL region}

We give the predictions for the cross section asymmetry and the polarization observables, for those models, described above, which give a good overall description of the available FFs data in SL and TL regions. The calculation is based on Eqs. (\ref{eq:asym}),  (\ref{eq:eq15}) and (\ref{eq:pol}), for a fixed value of the angle $\theta=\pi/4$.

\begin{figure}[pht]
\begin{center}
\includegraphics[width=16cm]{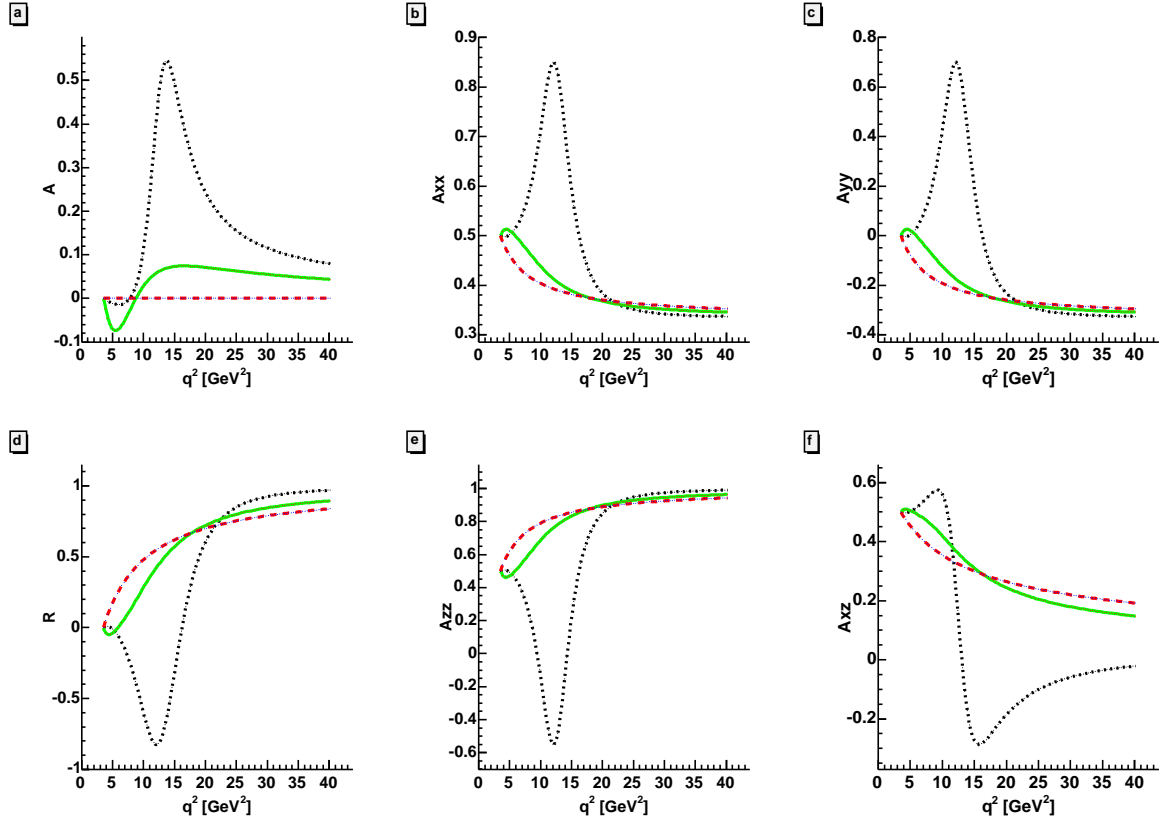}
\caption{\label{fig:fig3} Angular asymmetry and polarization observables, according to Eqs. (\protect\ref{eq:eq15}) and (\protect\ref{eq:pol}), for a fixed value of $\theta=45^0$. Notations as in Fig. \protect\ref{fig:fig2}.}
\end{center}
\end{figure}

As shown in Fig. \ref{fig:fig3}, all these observables are, generally, quite large. The model \cite{Ia73} predicts the largest (absolute) value at  $q^2\simeq$ 15 GeV $^2$ for all observables, except $A_{xz}$, which has two pronounced extrema.

All observables manifest a different behavior, according to the different models. The sign, also, can be opposite for VMD inspired models and pQCD. The model \cite{Lomon} is somehow intermediate between the two representations, as it contains the asymptotic predictions of QCD (at the expenses of a large number of parameters). 

The fact that single spin observables in annihilation reactions are discriminative towards models, especially at threshold, was already pointed out in Ref. \cite{Dub96}, for the process $e^++e^-\to p+\overline{p}$ on the basis of two versions of a unitary VDM model. The present results, (Fig. \ref{fig:fig3}), for the inverse reaction $p+\overline{p}\to e^++e^-$ confirm this trend and show that experimental data will be extremely useful, particularly in the kinematical region around $q^2\simeq$ 15 GeV $^2$. 

\section{Summary}

The measurement of polarization observables and the possibility to access individual nucleon FFs in TL and SL regions at larger $Q^2$ and/or with higher precision is foreseen in next future. 

A general analysis of the experimental data on nucleon electromagnetic FFs, extracted from elastic scattering and annihilation reactions, has been performed in the available kinematical region.

Expressions of the experimental observables in the reaction $p+\overline{p}\to e^++e^-$ have been derived in terms of the electromagnetic FFs, as a function of the momentum transfer squared. 

Some of the models on nucleon FFs have been reviewed, extended in TL region and used to give predictions on experimental observables which should be useful to plan future experiments.

Many questions are still open. Recent data in the SL region show that the ratio $G_{Ep}/G_{Mp}$  deviates from the expected dipole behavior. In the TL region, the values of $|G_M|$, obtained under the assumption  that $G_E=G_M$, are larger than the corresponding SL values. This has been considered as a proof of the non applicability of the Phr\`agmen-Lindel\"of theorem, (up to $s$=18 GeV$^2$, at least) or as an evidence that the asymptotic regime is not reached \cite{Bi93}.  The presence of a large relative phase of magnetic and electric proton FFs 
in 
the TL region, if experimentally proved at relatively large momentum transfer, 
can be considered a strong  
indication that these FFs have a different behavior. In particular, it will allow a test of the Phr\`agmen-Lindel\"of theorem \cite{Bi93}.

Large progress in view of a global interpretation of the nucleon FFs is expected from future experiments with antiproton beams: it will be possible, at the future FAIR  facility at GSI, to separate the electric and magnetic FFs in a wide region of $s$ and to extend the measurement of FFs up to the largest available energy, corresponding to $s\simeq 30$ GeV$^2$. 

The angular distribution of the produced leptons will allow the separation of the electric and magnetic FFs. The measurement of the asymmetry ${\cal R}$ (from the angular dependence of the 
differential cross section for $p+\overline{p}\leftrightarrow \ell^+ +\ell^-$) is sensitive to the relative value of $|G_M|$ and $|G_E|$.
In particular, the $\theta$-dependence of the single spin and double spin polarization observables is very sensitive to existing models of the nucleon FFs, which reproduce equally well the data in SL region.

Similar information can be obtained from the final polarization in $\ell^++\ell^- \to \vec p+\overline{p}$ \cite{Dub96}, but in this case one has to deal with the problem of hadron polarimetry, in conditions of very small cross sections.

Only the study of the processes $p+\overline{p}\to \pi^0+ \ell^+ +\ell^-$ and $p+\overline{p}\to \pi^++\pi^-+\ell^+ +\ell^-$, \cite{Re65,Dub95} will allow to measure proton FFs in the unphysical region (for $s\le 4m^2$, where the vector meson contribution plays an important role) and to determine the relative phase of pion and nucleon FFs.

\section{Appendix}

The Sachs FFs are expressed in terms of the Pauli and Dirac FFs as:

$$ G^N_{E}=F_1^N(Q^2)+\tau F_2^N(Q^2),~G^N_{M}=F_1^N(Q^2)+F_2^N(Q^2).$$
One can introduce the isoscalar and isovector FFs $F_i^{s}$ and $F_i^{v}$, $i=1,2$ as:
$2F^p_i=F_i^{s}+F_i^{v}$, $2F^n_i=F_i^{s}-F_i^{v}$. 

Then, the isoscalar and isovector currents can be parametrized in terms of meson propagators, 
effective FFs, and/or terms which insure specific properties, according to the different models.

\subsection{Model from Iachello, Jackson and Land\'e \protect\cite{Ia73} 
and Iachello and Wan \protect\cite{Wa04} }

FFs are parametrized following the work \cite{Ia73} , with a modification that consists in adding a phase in the dipole term, $g(Q^2)$, for the extension in TL region.
\begin{eqnarray*}
F_1^s(Q^2)&=&
\displaystyle\frac{g(Q^2)}{2}
\left[(1-\beta_\omega-\beta_\phi)+\beta_\omega\displaystyle\frac{\mu_\omega^2}{\mu_\omega^2+Q^2}+\beta_\phi
\displaystyle\frac{\mu_\phi^2}{\mu_\phi^2+Q^2}\right],\\
F_1^v(Q^2)&=&\displaystyle\frac{g(Q^2)}{2}
\left[(1-\beta_\rho)+\beta_\rho
\displaystyle\frac{\mu_\rho^2+8\Gamma_\rho\mu_\pi/\pi}
{(\mu_\rho^2+Q^2)+(4\mu_\pi^2+Q^2)\Gamma_\rho\alpha(Q^2)/\mu_\pi}\right],\\
F_2^s(Q^2)&=&
\displaystyle\frac{g(Q^2)}{2}
\left[(\mu_p+\mu_n-1-\alpha_\phi)
\displaystyle\frac{\mu_\omega^2}
{\mu_\omega^2+Q^2}+\alpha_\phi\displaystyle\frac{\mu_\phi^2}{\mu_\phi^2+Q^2}\right],\\
F_2^v(Q^2)&=&\displaystyle\frac{g(Q^2)}{2}
\left[(\mu_p-\mu_n-1)
\displaystyle\frac{\mu_\rho^2+8\Gamma_\rho\mu_\pi/\pi}{(\mu_\rho^2+Q^2)+(4\mu_\pi^2+Q^2)
\Gamma_\rho\alpha(Q^2)/\mu_\pi}\right],
\end{eqnarray*}
with $g(Q^2)=\displaystyle\frac{1}{(1+\gamma e^{i\theta}Q^2)^2}$ 
and $\alpha(Q^2)=\displaystyle\frac{2}{\pi}
\sqrt{\displaystyle\frac{Q^2+4\mu_\pi^2}{Q^2}}
ln\left[\displaystyle\frac{\sqrt{(Q^2+4\mu_\pi^2)}+\sqrt{Q^2}}{2\mu_\pi}\right]$, with the standard values of the masses $m=0.939$~GeV, $\mu_\rho=0.77$~GeV, $\mu_\omega=0.78$~GeV, $\mu_\phi=1.02$~GeV, $\mu_\pi=0.139$~GeV and the $\rho$ width $\Gamma_\rho=0.112$~GeV.

The values of the six parameters are given in Table \ref{table:IJL}.

\subsection{Model from Lomon \protect\cite{Lomon}}
\begin{eqnarray*}
F_1^{v}(Q^2)&=&
\displaystyle\frac{N}{2}
\left [
\displaystyle\frac{1.0317+0.0875(1+Q^2/0.3176)^{-2}}{(1+Q^2/0.5496)}+
\frac{g_{\rho'}}{f_{\rho'}}\displaystyle\frac{m_{\rho'}^2}{m_{\rho'}^2+Q^2}
\right ]
F_1^\rho(Q^2)+\\
&&\left(1-1.1192\displaystyle\frac{N}{2}-\displaystyle\frac{g_{\rho'}}{f_{\rho'}}\right)F_1^D(Q^2),\\
F_2^{v}(Q^2)&=&\displaystyle\frac{N}{2}
\left [\displaystyle\frac{5.7824+0.3907(1+Q^2/0.1422)^{-1}}{(1+Q^2/0.5362)}+
\kappa_{\rho'}\displaystyle\frac{g_{\rho'}}{f_{\rho'}}\displaystyle\frac{m_{\rho'}^2}{m_{\rho'}^2+Q^2}
\right ]
F_2^\rho(Q^2)+\\
&&
\left(\kappa_\nu-6.1731\displaystyle\frac{N}{2}-\kappa_{\rho'}\displaystyle\frac{g_{\rho'}}{f_{\rho'}}\right)F_2^D(Q^2),\\
F_1^{s}(Q^2)&=& \left (\displaystyle\frac{g_\omega}{f_\omega}\displaystyle\frac{m_{\omega}^2}{m_{\omega}^2+Q^2}+
\displaystyle\frac{g_{\omega '}}{f_{\omega '}}\displaystyle\frac{m_{\omega '}^2}{m_{\omega '}^2+Q^2}\right )
F_1^\omega(Q^2)+\\
&&\displaystyle\frac{g_\phi}{f_\phi}\displaystyle\frac{m_{\phi}^2}{m_{\phi}^2+Q^2}F_1^\phi(Q^2)+
\left(1-\displaystyle\frac{g_\omega}{f_\omega} -\displaystyle\frac{g_{\omega '}}{f_{\omega '}} \right)F_1^D(Q^2),\\
F_2^{s}(Q^2)&=& \left (\kappa_\omega\displaystyle\frac{g_\omega}{f_\omega}\displaystyle\frac{m_{\omega}^2}{m_{\omega}^2+Q^2}
+\kappa_{\omega '}\displaystyle\frac{g_{\omega '}}{f_{\omega '}}
\displaystyle\frac{m_{\omega '}^2}{m_{\omega '}^2+Q^2}\right )
F_2^\omega(Q^2)+\kappa_\phi\displaystyle\frac{g_\phi}{f_\phi}\displaystyle\frac{m_{\phi}^2}{m_{\phi}^2+Q^2}F_2^\phi(Q^2)+\\
&&
\left(\kappa_s-
\kappa_\omega\displaystyle\frac{g_\omega}{f_\omega}-
\kappa_{\omega '}\displaystyle\frac{g_{\omega '}}{f_{\omega '}}-
\kappa_\phi\displaystyle\frac{g_\phi}{f_\phi}\right)F_2^D(Q^2),\\
\end{eqnarray*}
with 
\begin{eqnarray*}
F_1^{\alpha,D}(Q^2)&=&\displaystyle\frac{\Lambda_{1,D}^2}{\Lambda_{1,D}^2+\widetilde Q^2}\displaystyle\frac{\Lambda_{2}^2}{\Lambda_{2}^2+\widetilde Q^2}, ~\alpha=\rho,~ \omega~ and~ \Lambda_{1,D}\equiv \Lambda_1~ \mbox{for} F_i^\alpha, ~\Lambda_{1,D}\equiv\Lambda_D ~for~F_i^D \\
F_2^{\alpha,D}(Q^2)&=&\displaystyle\frac{\Lambda_{1,D}^2}{\Lambda_{1,D}^2+\widetilde Q^2}\left(\displaystyle\frac{\Lambda_{2}^2}{\Lambda_{2}^2+\widetilde Q^2}\right)^2,~
F_1^\phi(Q^2)=F_1^\alpha\left(\displaystyle\frac{Q^2}{\Lambda_1^2+Q^2}\right)^{1.5},~\\
F_2^\phi(Q^2)&=&F_2^\alpha\left(\displaystyle\frac{\Lambda_1^2}{\mu_\phi^2}\displaystyle\frac{Q^2+\mu_\phi^2}{\Lambda_1^2+Q^2}\right)^{1.5} ,~
\widetilde Q^2=Q^2\displaystyle\frac{ln[(\Lambda_D^2+Q^2)/\Lambda_{QCD}^2]}{ln(\Lambda_D^2/\Lambda_{QCD}^2)}.
\end{eqnarray*}

The set of parameters is reported in Table \ref{table:lomon}.

\subsection{Model from Hohler \protect\cite{Ho76}}
This model is also based on a VMD parametrization:

\begin {eqnarray*}
F_1^\rho (Q^2)& = & 0.5 \left [0.955+\displaystyle\frac{0.09}{\left(1+{Q^2}/{0.355} \right)^2} \right] \displaystyle\frac{1}{1+{Q^2}/{0.536}},
\\
F_2^\rho (Q^2)& = & 0.5 \left [5.335+\displaystyle\frac{0.962}{\left(1+{Q^2}/{0.268} \right)^2}\right] \displaystyle\frac{1}{1+{Q^2}/{1.603}},
\\
F_{i}^{(s)}(Q^2) & = & \sum_j\displaystyle\frac{a_j^{(i,s)}}{b_j^{(s)}+Q^2},
\\
F_{i}^{(v)}(Q^2) & = & F_{i}^\rho(Q^2)+ \sum_j\displaystyle\frac{a_j^{(i,v)}}{b_j^{(v)}+Q^2}.
\end{eqnarray*}

The parameters are given in Table \ref{table:Hohler}.

\subsection{Model from Bosted \protect\cite{Bo95}}

The analytical expressions are inverse of polynomes as functions of $Q$,  whereas $G_{En}$ is described by a different function, as suggested by Galster \cite{Ga71}:
\begin{equation}
F^j=\displaystyle\frac{1}{1+\sum_i a^j_i Q^{i}},
\end{equation}

\begin{equation}
G_E^n=\displaystyle\frac{\alpha\mu_n\tau G_D(Q^2)}{1+\beta\tau },
\end{equation}
with $a^j_i$, $\alpha$ and $\beta$ free parameters. In the present notation $j=1,2,3$ corresponds to $G_{Ep}$ and $G_{Mn}$ and $G_{Mp}$, respectively.
The inverse polynomes are of fourth order for $G_{Ep}$ and $G_{Mn}$ and of fifth order for $G_{Mp}$.

The parameters are given in Table \ref{table:Bosted}.

{}

\begin{center}
\begin{table}[ht]
\begin{tabular}{|c|l|l|l|l|l|}
\hline
Reaction & $Q^2$ [GeV$^2]$& Observables& Laboratory  &Year&Reference \\
\hline
$e+p\to e+p$ &4.08-9.59 &$ G_{Mp}$& DESY&1966&Albrecht {\it et al.} \cite{Al65} \\
" &  0.16-0.85 & $G_{Ep}$,~ $G_{Mp}$&  SLAC & 1966& Janssens {\it et al.} \cite{Ja66}\\
" &  0.69 -  25.03 & $G_{Mp}$ &  SLAC & 1968    & Coward {\it et al.} \cite{Co68}\\
" &  0.4 - 2 & $G_{Ep}$,~$G_{Mp}$&  Bonn  &1971& Berger {\it et al.} \cite{Be71}\\
" &  0.670 -3.01& $G_{Ep}$,~ $G_{Mp}$& DESY &1973 &Bartel {\it et al.} \cite{Ba73}\\
* &  0.999-25.03&$ G_{Mp}$& SLAC & 1973& Kirk {\it et al.} \cite{Ki73}\\
" &  2.86-31.2 &$ G_{Mp}$ &SLAC& 1993   &Sill {\it et al.} \cite{Si92} \\ 
" &  1- 3 &$G_{Ep}$,~$G_{Mp}$&SLAC& 1994&  Walker {\it et al.} \cite{Wa93} \\
" &  1.75-8.83& $G_{Mp}$ &SLAC &1994& Andivahis {\it et al.}\cite{And94}\\
" &  0.65-5.20&$G_{Ep}$,~$G_{Mp}$&JLab, E94110 &2004  & Christy {\it et al.} \cite{Ch04}\\
" &  0.65-5.20&$G_{Ep}$,~$G_{Mp}$&JLab, Hall A &2004  & Qattan {\it et al.} \cite{Qa04}\\
$\vec e+p\to e+\vec p$& 0.49-3.47 &$G_{Ep}/G_{Mp}$&JLAB, Hall A&2000&Jones {\it et al.} \cite{Jo00}\\
$\vec e+p\to e+\vec p$& 3.5-5.5 &$G_{Ep}/G_{Mp}$&JLAB, Hall A&2002& Gayou {\it et al.} \cite{Ga02}\\
\hline
\end{tabular}
\caption[]{Data considered in the present analysis, for proton FFs, in SL region.} 
\label{table:prSL}
\end{table}
\clearpage\newpage

\end{center}
\begin{center}
\begin{table}[ht]
\begin{tabular}{|c|l|l|l|l|l|}
\hline
Reaction & $Q^2$ [GeV$^2$]& Observables& Laboratory  &Year&Reference \\
\hline
$ed\to epn$ &0.04- 1.16 &$G_{En}$ &SLAC& 1965 & Hughes {\it et al.} \cite{Hu65} \\
$ed\to epn$ &0.19,0.39,0.56 & $G_{Mn}$    & New York &1966 &Stein {\it et al.} \cite{St66} \\
$ed\to epn$ &0.39-0.565&$G_{Mn}$,~$G_{En}$&DESY   &1969&Bartel {\it et al.} \cite{Ba69}\\
$ed\to epn$ &0.28-1.8  &$G_{Mn}$          &Harvard&1973&Hanson {\it et al.} \cite{Ha73} \\
$ed\to epn$ &2.5-10    &$G_{Mn}$             &SLAC&1982& Rock {\it et al.} \cite{Ro82}\\
$^3\vec{He}$& 0.16&$G_{En}$& MIT &1991&Jones-Woodward {\it et al.} \cite{JW91} \\
$ed\to epn$  &0.435-1.36& $G_{Mn}$ &New York &1964& Akerlof {\it et al.} \cite{Ak64}\\ 
$\vec D (\vec e , e'\vec n)p $&.0.255& $G_{En}$&MIT &1994& Eden {\it et al.} \cite{Ed94} \\
$D(e,e'n)$,$D(e,e'p)$& 0.125-0.605&$G_{En}$ & Bonn& 1995&Bruins {\it et al.} \cite{Br95}\\
$D(e,e'n)$, $D(e,e'p)$&0.235-0.784 &$G_{En}$ &MAMI&1998& H. Anklin {\it et al.} \cite{An98}\\
$D(\vec e , e'\vec n)p$&0.15   & $G_{En}$    &MAMI&1999&Herberg {\it et al.} \cite{He99} \\
$ D (\vec e , e'\vec n)p$ &0.34 &$G_{En}$ & MAMI &1999& Ostrick {\it et al.} \cite{Os99}\\
$\vec D (\vec e , e'n)p$ &0.21&$G_{En}$   &NIKHEF&1999&Passchier {\it et al.} \cite{Pa99}\\
$^3\vec{He}(\vec e, e'n)pp$ &0.67&$G_{En}$   &MAMI  & 2003&  Bermuth {\it et al.} \cite{Be03} \\
$^3\vec{He} (\vec e , e')$ & 0.1-0.4 &$G_{En},G_{Mn},$& JLab & 2000&Golak {\it et al.}\cite{Go00}\\
$\vec D(\vec e , e'n)p$&0.495&$G_{En}$ &JLab&2001&Zhu {\it et al.} \cite{Zhu01}\\
$\vec D(\vec e , e'n)p$&0.5, 1&$G_{En}$ &JLab&2004&Warren {\it et al.} \cite{Day}\\
$D(\vec e , e'\vec n)p$&0.3-0.8&$G_{En}$ &MAMI&2004 &Glazier {\it et al.} \cite{Gl04}\\
$D(\vec e , e'\vec n)p$&0.5-1.5 & $G_{En}$&JLab&1999 &Madey {\it et al.} \cite{Madey}\\
\hline
\end{tabular}
\caption[]{Data considered in the present analysis, for neutron FFs,  in SL region.} 
\label{table:neutron}
\end{table}
\end{center}

\begin{center}
\begin{table}[ht]
\begin{tabular}{|c|c|c|c|l|}
\hline
Reaction & $q^2$ [GeV$^2$]& Laboratory& Year&Reference \\
\hline
$e^+e^-\to p\overline p$ &4.3      &ADONE, Frascati & 1973& Castellano {\it et al.} \cite{Ca73}\\
$ p\overline p\to e^+e^-$&3.52     &CERN            &1977 & Bassompierre {\it et al.} \cite{Ba77}\\
$p\overline p \to e^+e^-$&3.61     &CERN            &1983 &Bassompierre {\it et al.} \cite{Ba83}\\
$e^+e^-\to p\overline p$ &3.75-4.56& Orsay,DCI      &1979 & Delcourt {\it et al.} \cite{De79}\\ 
$e^+e^-\to p\overline p$ & 4.0-5.0  & Orsay, DCI    &1983&Bisello {\it et al.} \cite{Bi83}\\
$p\overline p\to e^+e^-$ & 8.9-13.0 & FERMILAB, E760&1993 &Armstrong {\it et al.} \cite{Ar92}\\
$p\overline p\to e^+e^-$ &3.52-4.18 &CERN, LEAR     &1994 & Bardin {\it et al.} \cite{Ba94}\\
$e^+e^-\to p\overline p$ & 3.69-5.95&ADONE, FENICE &1994 & Antonelli {\it et al.} \cite{An94}\\
$p \overline p \to e^+e^-$ &8.84 - 18.40 & FERMILAB, E835&1999 & Ambrogiani {\it et al.} \cite{Am99} \\ 
$p \overline p \to e^+e^-$ & 11.63- 18.22&FERMILAB, E835 &2003 &Andreotti {\it et al.}  \cite{An03}\\
\hline
$e^+e^-\to n\overline n$   &3.61- 5.95   & ADONE, FENICE& 1998 & Antonelli {\it et al.} \cite{An98}\\ 
\hline
\end{tabular}
\caption[]{Data considered in the present analysis for TL region.} 
\label{table:tl}
\end{table}
\end{center}

\begin{center}
\begin{table}[ht]
\begin{tabular}{|c|c|c|}
\hline
Parameters & GKex(02L) & This~work \\
\hline
$g_{\rho'}/f_{\rho'}$ &  0.0608  & -0.0152 \\
$\kappa_{\rho'}$      &  5.3038  & -30.315    \\
$g_\omega/f_\omega$   &  0.6896  &  0.5994  \\
$\kappa_\omega$       &  -2.8585 &  5.4557    \\
$g_\phi/f_\phi$       & -0.1852  &  -0.287304 \\
$\kappa_\phi$         & 13.0037  &  18.0208   \\
$\mu_\phi$            &  0.6848  &  0.802158  \\
$g(\omega ')/f_{\rho'}$ &  0.2346&  0.125397  \\
$k(\omega ')$         &  18.2284 &  15.4868   \\
$\Lambda_1$         &  0.9441    &  0.776982  \\
$\Lambda_D$         &  1.2350    &  1.06593   \\
$\Lambda_2$         &  2.8268    &  3.64885   \\
$\Lambda_{QCD}$     &  0.150     &  0.377101  \\
$N$                 &  1.        &  0.84501  \\ 
\hline
\end{tabular}
\caption[]{ Parameters of the model from \protect\cite{Lomon}. Fixed constants are $\kappa_v=3.706$, $\kappa_s=-0.12$, $m_\rho~=~0.776$~GeV, $m_\omega=0.784$~GeV, $m_\phi=1.019$~GeV, $m_{\rho'}=1.45$~GeV, $m_{\omega'}=1.419$~GeV.} 
\label{table:lomon}
\end{table}
\end{center}
\newpage\clearpage
\begin{center}
\begin{table}[ht]
\begin{tabular}{|c|c|c|}
\hline
Parameters & \protect\cite{Ia73,Wa04} & This~work \\
\hline
\hline
$\gamma$ &    0.250   &0.259   \\
$\beta_\rho$       &    0.672   &0.757  \\
$\beta_\omega$     &    1.102   &1.212 \\
$\beta_\phi$       &    0.112   &-0.114 \\
$\alpha_\phi$      &   -0.052   &-0.028 \\
$\theta$           &   $53^0$   &47$^0$  \\
\hline
\end{tabular}
\caption[]{ Parameters for the model from Ref. \protect\cite{Ia73,Wa04}.} 
\label{table:IJL}
\end{table}
\end{center}

\begin{center}
\begin{table}[ht]
\begin{tabular}{|c|c|c|c||c|c|c|c|}
\hline
FFs & par&\protect\cite{Ho76} & This~work &FFs & par&\protect\cite{Ho76} & This~work\\
\hline
\hline
$F_{is}$&$b_1^{(s)}$& 0.61   &0.49&    $F_{iv}$&$b_1^{(v)}$& 1.49 & 1.35  \\
        &$b_2^{(s)}$& 0.94   &0.84 &           &$b_2^{(v)}$& 4.33 &  3.22 \\
        &$b_3^{(s)}$& 3.20   &0.47&            &$b_3^{(v)}$& 8.47 & 21.28 \\
\hline
$F_{1s}$&$a_1^{(1,s)}$&  0.75  & 0.74& $F_{1v}$&$a_1^{(1,v)}$& 0.06 & 0.23  \\
        &$a_2^{(1,s)}$& -0.61  &-0.57&         &$a_2^{(1,v)}$&-0.32 & -0.55 \\      
        &$a_3^{(1,s)}$& -0.23  &-0.13&         &$a_3^{(1,v)}$& 0.10 & 0.04  \\        
\hline
$F_{2s}$&$a_1^{(2,s)}$& -0.15 &0.74 &  $F_{2v }$&$a_1^{(2,v)}$& -2.06  &-1.93\\
        &$a_2^{(2,s)}$& 0.18  &-0.57&	       &$a_2^{(2,v)}$&  0.23  &0.22\\
        &$a_3^{(2,s)}$& -0.03 &-0.13&	       &$a_3^{(2,v)}$&  0.23  &0.13\\
	\hline
\end{tabular}
\caption[]{ Parameters for the model from Ref. \protect\cite{Ho76}, corresponding to the fit in SL region, only. The starting set of parameters is from fit No. 9.1} 
\label{table:Hohler}
\end{table}
\end{center}

\begin{center}
\begin{table}[ht]
\begin{tabular}{|c|c|c|c|}
\hline
FF & $a_i^{(j)}$ & \protect\cite{Bo95} & This~work \\
\hline
\hline
$G_{Ep}$&$a_1^1$&0.62 &-1.55 \\ 
        &$a_2^1$&0.68 &10.40\\
        &$a_3^1$&2.80 &-11.53 \\ 
        &$a_4^1$&0.83 &7.87\\
\hline
$G_{Mp}/\mu_p$&$a_1^2$&0.35&   0.28  \\  
              &$a_2^2$&2.44 &2.61  \\ 
              &$a_3^2$&1.04 &0.41\\ 
              &$a_4^2$&0.34 &1.03 \\
              &$a_5^2$&0.34 &0.34 \\
\hline	      
$G_{Mn}/\mu_n$&$a_1^3$&-1.74& 0.14 \\
              &$a_2^3$&9.29 &  3.39 \\
              &$a_3^3$&-7.63 &-2.07 \\ 
              &$a_4^3$&4.63 &3.09 \\
\hline	      			   
$G_{En}$&$\alpha$&  1.25&0.43\\
        &$\beta$& 18.3 &-0.22 \\
\hline
\end{tabular}
\caption[]{ Parameters for the model from Ref. \protect\cite{Bo95}, corresponding to the fit in SL region, only.} 
\label{table:Bosted}
\end{table}
\end{center}
\end{document}